\documentclass[pre,twocolumn,superscriptaddress,floatfix]{revtex4-1}  

\usepackage[paperwidth=210mm,paperheight=297mm,centering,hmargin=2cm,vmargin=2.5cm]{geometry}

\usepackage[pdftex]{graphicx}
\usepackage{amsmath,amsfonts,amssymb}
\usepackage{natbib}
\usepackage{MnSymbol}

\usepackage{hyperref}
\usepackage{xcolor} 
\usepackage{array}
\usepackage{pgfplots}
\usepackage{graphicx}
\usepackage[left,modulo,pagewise]{lineno}
\usepackage{dsfont}

\definecolor{bluemoi}{rgb}{0.25,0.50 ,0.75} 

\hypersetup{
  bookmarksopen=true,
  pdftitle=" ",
  pdfauthor=" ", 
  pdfsubject=" ", 
  pdftoolbar=true, 
  pdfmenubar=true, 
  pdfhighlight=/O, 
  colorlinks=true, 
  pdfpagemode=UseNone, 
  pdfpagelayout=SinglePage, 
  pdffitwindow=true, 
  linkcolor=bluemoi, 
  citecolor=bluemoi, 
  urlcolor=bluemoi 
}

\newcommand{\titre}{A generic framework for the development of geospatial processing pipelines on clusters}

\makeatletter
\renewcommand{\figurename}{\sf \textbf{Figure}}
\renewcommand{\thefigure}{\arabic{figure}}
\renewcommand{\fnum@figure}{\sf\textbf{\figurename}~\textbf{\thefigure}}
\renewcommand{\tablename}{\sf\textbf{Table}}
\renewcommand{\thetable}{\arabic{table}}
\renewcommand{\fnum@table}{\sf\textbf{\tablename}~\textbf{\thetable}}
\makeatother

\begin{document}

\title{\titre}
\author{R\'emi~Cresson~and~Gabriel~Hautreux}
\thanks{Corresponding author: remi.cresson@irstea.fr}

\begin{abstract}
The amount of remote sensing data available to applications is constantly growing due to the rise of very-high-resolution sensors and short repeat cycle satellites. Consequently, tackling computational complexity in Earth Observation information extraction is rising as a major challenge. Resorting to High Performance Computing (HPC) is becoming a common practice, since it provides environments and programming facilities able to speed-up processes. In particular, clusters are flexible, cost-effective systems able to perform data-intensive tasks ideally fulfilling any computational requirement. However, their use typically implies a significant coding effort to build proper implementations of specific processing pipelines.
This paper presents a generic framework for the development of RS images processing applications targeting cluster computing. It is based on common open sources libraries, and leverages the parallelization of a wide variety of image processing pipelines in a transparent way. Performances on typical RS tasks implemented using the proposed framework demonstrate a great potential for the effective and timely processing of large amount of data. 
\end{abstract}

\maketitle

\section{Introduction}
There is an increasing volume of remote sensing (RS) images for earth observation in recent years. Satellite and airborne RS data become widespread and sensors are currently sustaining major technical revolution. In one hand, the use of very high resolution RS data is booming, and both spectral and spatial resolutions are each generation sharper \cite{BEN}. In the other hand, there is a growing need of sensors with close temporal acquisition in Earth Sciences. With the arrival of satellites such as Sentinel constellation (10m spatial resolution, 5 days revisit cycle), we entered a new era of earth observation. High orbital revisit frequency allow to monitor in near real time the earth's surface, which strongly benefits to many sectors, e.g. agricultural~\cite{LOB}. 
Currently, Earth Observation is producing permanently a stream of data. 
Therefore, extracting information from existing RS data became a major computational challenge~\cite{DOR}. 

For this purpose, high performance computing (HPC) techniques provide solutions to speed-up computations allowing the processing of large data volumes in reasonable time~\cite{PLA}. 
Basically, it relies on parallelization to increase computational power~\cite{ASA}, and involves various computing environments and programming facilities. 
Several approaches of HPC have been used with RS data, like multiprocessors, computers networks (e.g. clusters, grids, clouds), and hardware such as graphics processing units (GPU).
While multiprocessors and GPUs work usually in a shared memory context, i.e. on the same data stored in memory, clusters and clouds are distributed systems, aiming to handle distinct parts of the data bulk. Clusters are homogeneous systems (i.e. similar hardware and software) composed of tightly coupled machines (e.g. Thunderhead Beowulf cluster at NASA's Goddard Space Flight Center), whereas grids and clouds are heterogeneous and loosely coupled systems, generally on a larger scale~\cite{BER}. It can also be noted that grids and clouds can be composed of several clusters, and most clusters exploit multiprocessors and GPUs as co-processors. Hence, HPC techniques are handled in a complementary pattern. Although each one may suit to a particular application, clusters are considered by the Earth and space sciences community as a cost-effective system able to satisfy specific computational requirements~\cite{LEE}. This is why in this paper, we choose to focus on cluster based systems only. In the following we call "cluster" a group of machines with the same hardware, namely "nodes". Nodes are connected by a fast local area network, sharing a parallel file storage system.

In this context, one crucial point is the software implementation. 
Often the literature presents algorithms adapted to HPC architectures for one very specific task. Therefore, coding is not generic and
 each new algorithm implementation require expertise, often at considerable cost. We can distinguish common parallel programing paradigms for various HPC architectures. OpenMP and Pthreads are commonly used to carry out parallel computations with multiple processors in a shared memory environment. For writing programs on GPUs, CUDA and OpenCL frameworks are frequently used. In cluster based systems, the Message Passing Interface (MPI) programming model is commonly used to manage communications 
between nodes. Besides, hybrid parallel algorithms are regularly employed to achieve the best performances, e.g. MPI + CUDA on a GPU cluster. Hence, developing RS processing pipelines on HPC architectures require some advanced knowledge of both hardware and programming paradigms, holding back its democratization especially for research and academics. 
Currently, existing popular libraries have programming interfaces embedding multiprocessing in a shared memory environment~\cite{ITK} and GPUs~\cite{CHR}. There is only a few studies presenting frameworks for RS image processing on clusters, mainly relying on MPI~\cite{GUA2} and hybrid approaches~\cite{QIN2}. However, up to our knowledge, there is no available cluster-oriented paradigm benefiting from multiprocessing in a shared memory environment and GPUs.
In this paper, we present a generic framework for the development of geospatial processing pipelines on clusters, which is (i) open-source and portable (cross-platform), (ii) developer friendly, with strong abstraction of low level cluster related mechanisms, (iii) based on an existing rich image processing library, Orfeo Toolbox (OTB,~\cite{OTB}) and relying on the MPI standard. Our approach benefits from multiprocessing in a shared memory environment as well as GPU support, while ensuring the distribution of the entire data across clusters. We first give in Section \ref{sec_method} a detailed description of our cluster-oriented RS image processing parallelization framework. Our approach is then successfully tested with a set of popular pipelines on some Spot 6 satellite images (Section \ref{sec_data}). We finally discuss the main advantages and limits of our approach in Section \ref{sec_disc}.

\section{Method}
\label{sec_method}

\subsection{Overview}
\label{sssec_Overview}

We consider the library for RS image processing, OTB, built on top of an application development framework widely used in medical image processing (ITK, the Insight Toolkit~\cite{ITK}), and a cluster composed of several nodes with a parallel file system. Our goal is to bring a development framework to exploit the cluster processing capabilities.
The parallelization of already implemented pipelines should be enabled with the minimum effort, and the opportunity to build clusters compliant pipelines must be granted to non-expert developers (namely users) without fully understanding the low level MPI implementation. The existing support for multiprocessing in a shared memory environment and GPU brought by ITK and OTB must also be preserved.
In the following sections, we provide description of the existing image processing framework of the used libraries (Section \ref{ss_Pipelines}). Then we introduce the concepts of parallel process objects in Section \ref{sss_parallel_process_objects} and parallelized pipeline in Section \ref{sss_parallelized_pipeline}. Finally, we detail our solution to overcome the problem of geospatial raster data output in Section \ref{ss_writer}.

\subsection{Description of a pipeline}
\label{ss_Pipelines}

This section describes the terminology and the execution steps of a RS processing pipeline, from the OTB perspective. A pipeline is a directed graph of process objects, that can be: 
\begin{itemize}
\item Sources: initiating the pipeline. Generating input data objects (e.g. image file reader),
\item Filters: processing the data objects,
\item Mappers: terminating the pipeline. Writing data on disk (e.g. image file writer), or interfacing with some other system (e.g. display).
\end{itemize}
Sources and filters can generate one or multiple data objects (e.g. image, mesh, vector, matrix, number). For the user, building a pipeline consists in connecting process objects together. The execution of a pipeline starts by triggering process objects, usually mappers. More details about these objects and architecture are provided in~\cite{OSA}.

The architecture of the libraries hides the complexity of internal mechanisms for pipeline execution, which involve several important steps. First of all, process objects that need to be triggered are determined, to avoid redundant execution.  

Then, the execution of the pipeline is started by a mapper trigger. When a filter or a mapper is triggered, a signal is sent upstream to request information about mandatory input data (i.e. information about output data of upstream process object(s)). In this way, it is propagated back through the pipeline, from mappers to sources via filters. Once this request reach sources, information are generated from metadatas. This information can be image size and pixel spacing, that are propagated downstream to mapper. It may be noted that filters can potentially modify these information, according to the process they implement (e.g. resampling might change output image size). Finally, they reach the mapper, initiating the data processing. 

Information regarding the size of the image that has to be produced, is then used by the mapper to choose a splitting strategy. Typically, the splitting scheme is based on the system memory specification. Other strategies can be chosen, e.g. striped or tiled regions with fixed dimensions. Once the splitting strategy is determined, the mapper proceed, typically write on disk. The mapper process the image sequentially, region by region, requesting its input filter. 
The data request and generation is handled through the pipeline in the same way as for 
the information: once the request reaches the sources, initiating the pipeline, the requested region is produced, then processed through filters, to finally end in the mapper. The pipeline execution continues with the next image region.

\subsection{Toward parallel process objects}
\label{sssec_Parallel}

The idea is to go through a cluster-oriented parallel approach, while preserving the existing development framework of the libraries. This includes the coding of process object (sources, filters, mappers), and pipelines creation on a higher abstraction level.
In the following, we detail two concepts. We first introduce parallel process objects, which are basically process objects implementing a MPI based abstraction layer. Second, we present the parallelized pipeline, which designates the set of pipelines running across the cluster, each one being a different MPI process.

\subsubsection{Parallel process objects}
\label{sss_parallel_process_objects}
Regarding the parallel approach, two kind of process objects generating images must be distinguished. The first ones can produce output images in a region independent fashion, meaning that identical pixels are generated whatever the output requested images region. 
Hence, an entire output image can be gathered from multiple generation of different regions, making these process objects straightforwardly suitable for parallel approach. The second kind of process objects generates output images that are dependent of the requested region and, thus, require specific implementation. However, the ITK and OTB libraries propose a convenient development framework to greatly reduce the developer task in implementing such algorithms. This is particularly useful to design and implement filters which, for instance, process data using multi-process on shared memory (namely \textit{Multithreaded} filters) and filters that persist data through multiple update (namely \textit{Persistent} filters). The advantage of \textit{Multithreaded} filters is that the developer does not need to be an expert in low level threads management, neither knowing the end user's hardware specifications. Typically, he just has
  to implement some specific methods. These methods are generally data generation on thread region (\textit{ThreadedGenerateData}), and shared resources handling before and/or after the multi-threaded part (e.g. \textit{Before/AfterThreadedGenerateData}). Similarly, \textit{Persistent} filters implement methods to handle variables needing to be updated during the process, e.g. pixels statistics (\textit{Synthesis}, \textit{Reset}). 
In their parallelized version, such filters must induce communications between MPI processes to aggregate variables over the cluster. This is achieved using the MPI with \textit{many-to-one}, \textit{one-to-many} or \textit{many-to-many} communication patterns in the previously mentioned methods. 

\subsubsection{Parallelized pipeline}
\label{sss_parallelized_pipeline}

As described in the previous part, libraries offer an abstraction layer for the low level multiprocessing paradigm. In addition, a wide range of existing filters use GPU. 
To take benefit from these advantageous supports, working only in a shared memory context, we focus on the MPI processes grain level to parallelize pipelines: we distribute one pipeline per MPI process, ensuring the shared memory context, then pipelines are executed simultaneously to work in a collaborative fashion.
A parallelized pipeline can be composed of parallelized or native process objects, depending if the implemented algorithm produces the same unique result whatever the requested region (as explained in Section \ref{sss_parallel_process_objects}).
Given the typology of our parallelized pipeline, it must be terminated with a parallel mapper that ensures the load balancing of the cluster. This is achieved by computing a splitting scheme as described in Section \ref{ss_Pipelines}, and determining the way of distributing data across the cluster.

\subsection{Raster writing on parallel file systems}
\label{ss_writer}

Most of RS processing applications produce images in raster format. 
Unfortunately, this kind of data have large size, leading to an I/O (Input/Output) bottleneck~\cite{WAN}. 
For this purpose, we develop a mapper able to write GeoTiff file on parallel file systems. 
Thanks to our implementation using MPI-IO (MPI subroutines for file access~\cite{THA}), multiple MPI processes can write their piece of data simultaneously in the same unique file. 
We chose to write files in the GeoTiff format, in a row-wise, interleaved pixel fashion, which is faster than tile-wise~\cite{QIN}. 
Our writer implements various strategies for cluster-oriented splitting scheme described in Section \ref{ss_Pipelines}. Either size or number of splits can be manually set, or automatically computed using the system specifications (memory and number of MPI processes). Our writer has a static load balancing, meaning that each process has a fixed processing schedule.

\section{Experiments}
\label{sec_data}
The method presented in Section \ref{sec_method} is implemented in C++ using OTB, ITK and MPI. We parallelized large number of already implemented pipelines in OTB~\cite{CHR2}. In this section, we first analyze performances in reading and writing images (Section \ref{bench_IO}), then we present the results related to the speedup of a set of pipelines frequently used in RS image processing (Section \ref{bench_Pipelines}). To conduct this series of experiments, we used a dataset of very-high-resolution Spot 6 images acquired within the GEOSUD project\footnote{www.equipex-geosud.fr} presented in Table \ref{tab_data}.
\begin{table}[!t]
\renewcommand{\arraystretch}{1.3}
\caption{Characteristics of the dataset}
\label{tab_data}
\centering
\begin{tabular}{|l|l|l|l|l|}
\hline
Id & Product type & \begin{tabular}{@{}c@{}}Image size \\ ($col \times row \times band$)\end{tabular}  & \begin{tabular}{@{}c@{}}Values \\ encoding\end{tabular} & \begin{tabular}{@{}c@{}}File \\ size\end{tabular}\\
\hline
\textit{XS} & Multispectral & $10699 \times 11899 \times 4$ & 16 bits & 1.0 Gb\\
\textit{PAN}  & Panchromatic  & $42599 \times 47299 \times 1$ & 16 bits & 4.0 Gb\\
\hline
\end{tabular}
\end{table}
All experiments are conducted on a cluster composed of 16 DELL R630 Intel Xeon E5-2690 nodes, each one made of 2 sockets of 12 cpus at 2,6Ghz and 64 Gb RAM. Machines are linked with a very fast network (Infiniband 40 Gb/s). All files were stored on a RAID disk array shared to the cluster with a General Parallel File System. 
We binded MPI process with socket rather than node, to populate each socket exclusively by threads from one unique MPI process.
This ensures optimal caching, enabling efficiency of the ITK and OTB multi processing framework which rely on shared memory. Therefore, each node hosts two MPI processes.

\subsection{I/O performances}
\label{bench_IO}

We used a simple parallel pipeline, composed of a source and our parallel writer, for measuring I/O related durations. 

\begin{figure}[!t]
\centering
\includegraphics[width=3in]{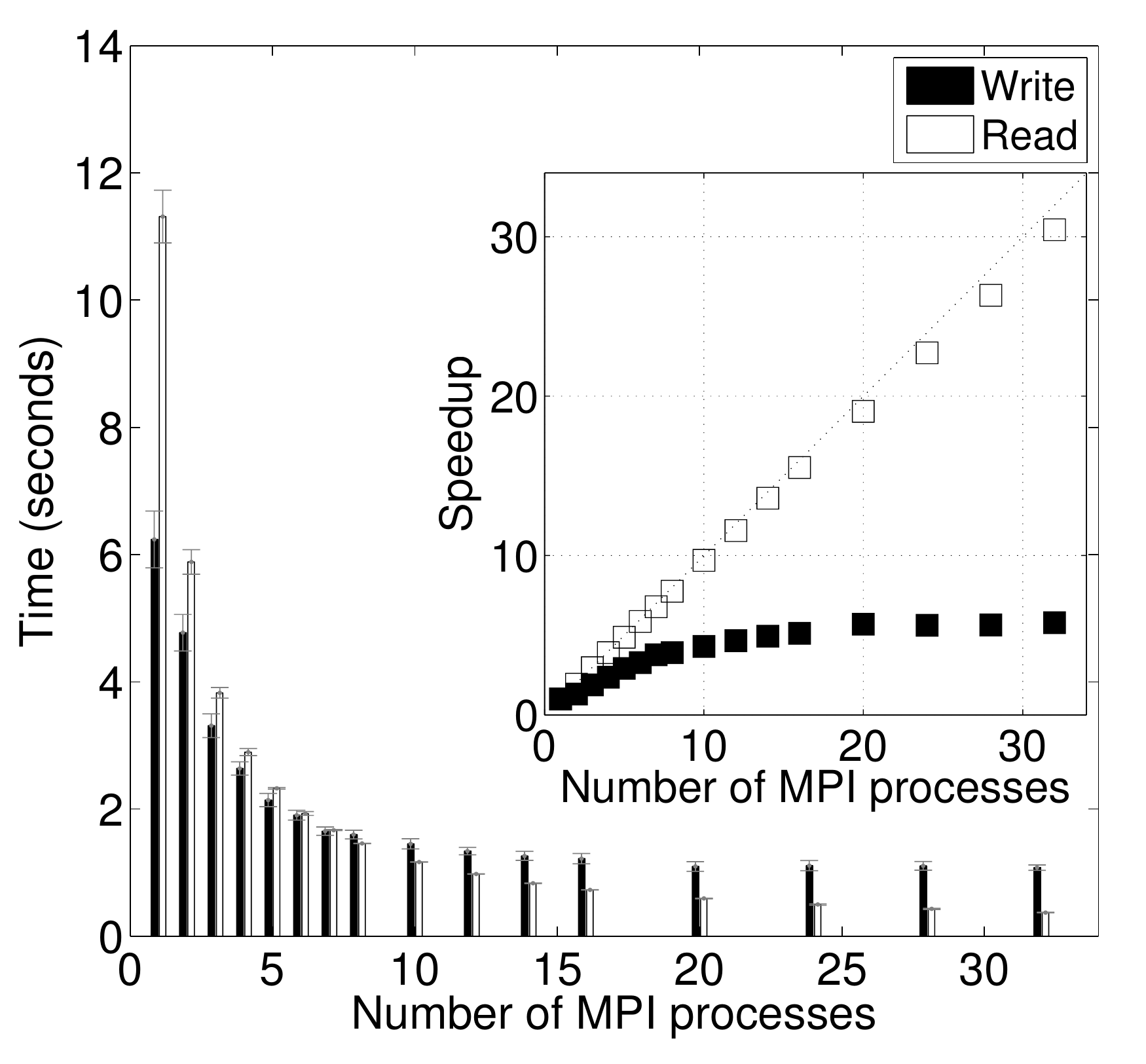}
\caption{Time consumed by reading (white bars) and writing (black bars) the Spot 6 GeoTiff image stored in the cluster file system, with error bars. An inset presents the speedup of reading (white boxes) and writing (black boxes).} 
\label{fig_io_bench}
\end{figure}

Figure \ref{fig_io_bench} shows measures dedicated to reading and writing operations from the \textit{XS} GeoTiff image file. 

Time spent in reading decreases linearly with the number of MPI processes. 
This operation is quite scalable, with a speedup close to the number of processes. Yet, the speedup for writing is not linear and reaches a maximum of 6.1 for 32 processes. This poor scalability is related to the striping for writing data, which needs to be fine tuned to fit better the file system use. Besides, writing speedup is expected to be lower than reading because more processes synchronization is required.

\subsection{Parallelized pipelines}
\label{bench_Pipelines}
We derived the following OTB pipelines $P_i$ in their parallelized flavor, replacing each original image file writer by our parallel one:
\begin{description}
\item[($P_1$) Orthorectification]
\item[($P_2$) Textures Extraction, using Haralick indicators]
\item[($P_3$) Pansharpening, from \textit{PAN} and \textit{XS} images]
\item[($P_4$) Image classification, using a random forest based rule]
\item[($P_5$) Meanshift filtering]
\item[($P_6$) Conversion from Jpeg2000 format to GeoTiff]
\item[($P_7$) Resampling \textit{XS} image over \textit{PAN} image]
\item[(\textit{I/O}) Reading and writing the image in another file]
\end{description}
To analyze the computational performance of our approach, we decompose run time of each $P_i$ for several number of MPI processes, and calculate the corresponding speedup.

\begin{table}[!t]
\renewcommand{\arraystretch}{1.15}

\caption{Run times and speedup of pipelines for $N$ MPI processes}
\label{tab_result}

\begin{tabular}{|l|l|l|l|l|}
\hline
\begin{tabular}{@{}c@{}} 
N
\end{tabular} & 1 & 2 & 4 & 8 \\
\hline

$P1$ & 1911s  & 970s ($\times$\textbf{2.0}) & 481s ($\times$\textbf{4.0}) & 241s ($\times$\textbf{7.9}) \\
 & \tiny {$\pm$17s} & \tiny {$\pm$9.3s (0.02)} & \tiny {$\pm$4.0s (0.03)} & \tiny {$\pm$1.6s (0.05)}  \\
$P2$ & 498s  & 260s ($\times$\textbf{1.9}) & 125s ($\times$\textbf{4.0}) & 64s ($\times$\textbf{7.8})  \\
 & \tiny {$\pm$13s} & \tiny {$\pm$8.6s (0.06)} & \tiny {$\pm$0.48s (0.02)} & \tiny {$\pm$2.0s (0.24)}  \\
$P3$ & 2613s  & 1339s ($\times$\textbf{2.0}) & 674s ($\times$\textbf{3.9}) & 347s ($\times$\textbf{7.5}) \\
 & \tiny {$\pm$18s} & \tiny {$\pm$13s (0.02)} & \tiny {$\pm$2.3s (0.01)} & \tiny {$\pm$3.0s (0.06)} \\
$P4$ & 495s  & 471s ($\times$\textbf{1.1}) & 127s ($\times$\textbf{3.9}) & 63s ($\times$\textbf{7.9}) \\
 & \tiny {$\pm$2.2s} & \tiny {$\pm$1.7s (0.00)} & \tiny {$\pm$0.36s (0.01)} & \tiny {$\pm$0.16s (0.02)} \\
$P5$ & 1972s  & 851s ($\times$\textbf{2.4}) & 522s ($\times$\textbf{3.9}) & 284s ($\times$\textbf{7.1}) \\
 & \tiny {$\pm$137s} & \tiny {$\pm$113s (0.26)} & \tiny {$\pm$85s (0.61)} & \tiny {$\pm$41s (0.98)} \\
$P6$ & 1013s  & 526s ($\times$\textbf{1.9}) & 262s ($\times$\textbf{3.9}) & 127s ($\times$\textbf{8.0}) \\
 & \tiny {$\pm$2.5s} & \tiny {$\pm$4.6s (0.02)} & \tiny {$\pm$1.4s (0.02)} & \tiny {$\pm$0.21s (0.01)} \\
$P7$ & 2192s  & 1106s ($\times$\textbf{2.0}) & 561s ($\times$\textbf{3.9}) & 285s ($\times$\textbf{7.7}) \\
 & \tiny {$\pm$16s} & \tiny {$\pm$11s (0.02)} & \tiny {$\pm$1.2s (0.01)} & \tiny {$\pm$0.82s (0.02)} \\
$IO$ & 17s  & 11s ($\times$\textbf{1.6}) & 5.5s ($\times$\textbf{3.2}) & 3.0s ($\times$\textbf{5.7}) \\
 & \tiny {$\pm$0.32s} & \tiny {$\pm$0.16s (0.02)} & \tiny {$\pm$0.07s (0.04)} & \tiny {$\pm$0.08s (0.15)} \\

\hline
\end{tabular}

\begin{tabular}{|l|l|l|l|l|}
\hline
\begin{tabular}{@{}c@{}} 
\small{N}
\end{tabular} & 12 & 16 & 32\\
\hline

$P1$ & 163s ($\times$\textbf{12}) & 125s ($\times$\textbf{15}) & 63s ($\times$\textbf{31}) \\
 & \tiny {$\pm$1.2s (0.08)} & \tiny {$\pm$0.36s (0.04)} & \tiny {$\pm$0.04s (0.02)} \\
$P2$ & 42s ($\times$\textbf{12}) & 31s ($\times$\textbf{16}) & 15s ($\times$\textbf{32}) \\
 & \tiny {$\pm$1.00s (0.27)} & \tiny {$\pm$0.81s (0.41)} & \tiny {$\pm$0.42s (0.85)} \\
$P3$ & 241s ($\times$\textbf{11}) & 188s ($\times$\textbf{14}) & 115s ($\times$\textbf{23}) \\
 & \tiny {$\pm$2.0s (0.09)} & \tiny {$\pm$0.56s (0.04)} & \tiny {$\pm$5.2s (0.96)} \\
$P4$ & 44s ($\times$\textbf{11}) & 34s ($\times$\textbf{15}) & 17s ($\times$\textbf{29}) \\
 & \tiny {$\pm$0.44s (0.12)} & \tiny {$\pm$0.31s (0.13)} & \tiny {$\pm$0.12s (0.21)} \\
$P5$ & 186s ($\times$\textbf{11}) & 152s ($\times$\textbf{13}) & 69s ($\times$\textbf{29}) \\
 & \tiny {$\pm$19s (1.1)} & \tiny {$\pm$15s (1.4)} & \tiny {$\pm$3.6s (1.5)} \\
$P6$ & 94s ($\times$\textbf{11}) & 64s ($\times$\textbf{16}) & 33s ($\times$\textbf{31}) \\
 & \tiny {$\pm$0.99s (0.11)} & \tiny {$\pm$0.15s (0.04)} & \tiny {$\pm$0.08s (0.07)} \\
$P7$ & 196s ($\times$\textbf{11}) & 151s ($\times$\textbf{14}) & 84s ($\times$\textbf{26}) \\
 & \tiny {$\pm$0.70s (0.04)} & \tiny {$\pm$0.73s (0.07)} & \tiny {$\pm$0.57s (0.18)} \\
$IO$ & 2.3s ($\times$\textbf{7.6}) & 1.9s ($\times$\textbf{9.0}) & 1.5s ($\times$\textbf{12}) \\
 & \tiny {$\pm$0.06s (0.21)} & \tiny {$\pm$0.06s (0.29)} & \tiny {$\pm$0.07s (0.51)} \\

\hline
\end{tabular}

\end{table}
  
\begin{figure}[!t]
\centering
\includegraphics[width=3.5in]{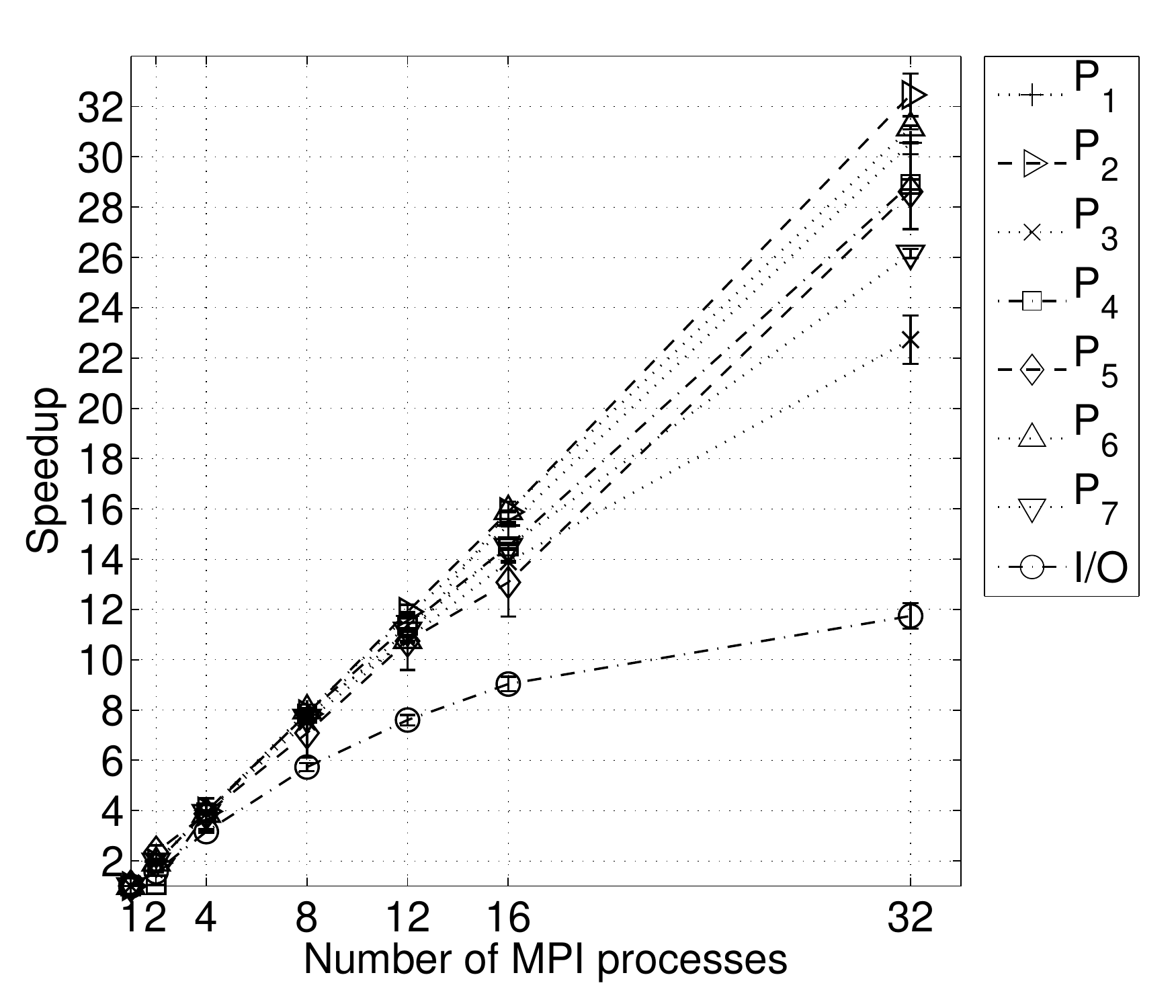}
\caption{Measured speedup of $P_i$ for increasing number of MPI processes.}
\label{fig_pi_bench}
\end{figure}

Table~\ref{tab_result} presents measured run times in seconds and Figure~\ref{fig_pi_bench} shows the speedup.

The speedup is more or less equal to the number of MPI processes, characterizing a good scalability, except for the single I/O operation. 
Besides, Section \ref{bench_IO} shows that writing operation limits the scalability, explaining why the speedup is slightly below the number of MPI processes for each $P_i$. 

\section{Discussion}
\label{sec_disc}

\subsection{A novel approach to create cluster oriented pipelines}
Our first goal was to provide a framework to develop cluster oriented applications for RS images processing. The solution we propose takes roots 
in the extensively used open-source libraries ITK and OTB. The original development framework was enriched to embed the MPI, for inter-nodes communication and parallel file access. 
We introduced  parallel process objects, and parallel pipelines, enabling the execution of one pipeline on multiple nodes of a cluster.

\subsection{Scalability}
Our research also sought to result in operational and scalable processes. The approach was successfully applied to a set of common RS pipelines, on Spot 6 imagery at very-high-resolution. Run times were measured for various number of MPI processes, and demonstrate a good scalability of tested parallel pipelines. In the near future, it will be used to extract a wide range of information for earth observation, on a daily basis from satellite imagery acquired through the GEOSUD platform and the THEIA Land Data Center\footnote{http://www.theia-land.fr/en}. 

\subsection{Limits and further research}
It should be noted that the presented method also has limits, which may restrict its use in some cases. 
To start with, we used a small cluster which has only a dozen of nodes to lead our tests. Hence, we could not fully investigate the limits of our approach in term of scalability on many nodes.
Besides, a key drawback of the use of clusters is the I/O bottleneck, which is common in HPC. 
In our experiments, we show that the scalability of one pipeline is closely dependent on the balance between processing time and file access time.
While pipelines are generally scalable, I/O operations might require fine tuning according to the file system specifications.
Nevertheless, a cluster should not be used to speed up a process spending a lot more time in writing data than processing. 
Another possible improvement would be to use a dynamic load balancing, which might tackle the problem of algorithms running in a non-constant time on different image regions.
Finally, our approach is only suitable for filters composing a pipeline that could be parallelized using communications between nodes. Depending on the implemented algorithm, some filters might not fulfill this condition, or could lead to a poor speedup.  
Regarding this last issue, we recommend to split this kind of pipeline in multiple homogeneous parts with uniform scalability
 and to run them sequentially. This puts forward the question related to the orchestration of multiple connected pipelines execution, which should be addressed in the future.

\section{Conclusion}
This work was carried out with a view toward processing very-high-resolution and high-acquisition-rate satellite images on clusters. We propose a cluster oriented framework for the development of remote sensing images processing applications, using the Orfeo Toolbox and the Message Passing Interface. We parallelized successfully a number of existing pipelines, and demonstrated the good scalability of the processes. Parallel pipelines will be executed every day to extract Earth observation data from very-high-resolution and multi-temporal  images acquired through the GEOSUD platform and the THEIA Land Data Center, at CNES (the french Space Agency) and Irstea (the french Research Institute of Science and Technology for Environment and Agriculture). Further research could focus on improving the load balancing strategies, as well as the orchestration of multiple connected pipelines execution. The source code corresponding to the pipelines presented in this paper is available for download at~\cite{GIT}, and the exposed framework will be integrated in the forthcoming releases of the Orfeo Toolbox\footnote{http://www.orfeo-toolbox.org}.

\section*{Acknowledgments}
This work was supported by public funds received through GEOSUD, a project (ANR-10-EQPX-20) of the \textit{Investissements d'Avenir} program managed by the French National Research Agency.
The authors would like to thank the reviewers for their valuable comments and suggestions which have greatly contributed in improving the manuscript. 
The authors would also like to thank Maxime Lenormand, Raffaele Gaetano and Julien Michel for their great help, and the OTB community for the precious support. They also thank the CINES supercomputing center and the HPC@LR competence center for providing the HPC support.

\bibliographystyle{unsrt}
\bibliography{cresson2016_final}

\begin{thebibliography}{10}

\bibitem{BEN}
J.~A. Benediktsson, J.~Chanussot, and W.~M. Moon.
\newblock Advances in very-high-resolution remote sensing.
\newblock {\em Proceedings of the IEEE}, 101, 2013.

\bibitem{LOB}
D.~B. Lobell and G.~P. Asner.
\newblock Cropland distribution from temporal unmixing of modis data.
\newblock {\em Remote Sensing of Environment}, 93:412--422, 2004.

\bibitem{DOR}
J.~Dorband, J.~Palencia, and U.~Ranawake.
\newblock Commodity computing clusters at goddard space flight center.
\newblock {\em J. Space Commun.}, 3:1, 2003.

\bibitem{PLA}
Antonio~J Plaza and Chein-I Chang.
\newblock {\em High performance computing in remote sensing}.
\newblock CRC Press, 2007.

\bibitem{ASA}
Krste Asanovic, Ras Bodik, Bryan~Christopher Catanzaro, Joseph~James Gebis,
  Parry Husbands, Kurt Keutzer, David~A Patterson, William~Lester Plishker,
  John Shalf, Samuel~Webb Williams, et~al.
\newblock The landscape of parallel computing research: A view from berkeley.
\newblock Technical report, Technical Report UCB/EECS-2006-183, EECS
  Department, University of California, Berkeley, 2006.

\bibitem{BER}
Armando Fox, Rean Griffith, Anthony Joseph, Randy Katz, Andrew Konwinski, Gunho
  Lee, David Patterson, Ariel Rabkin, and Ion Stoica.
\newblock Above the clouds: A berkeley view of cloud computing.
\newblock {\em Dept. Electrical Eng. and Comput. Sciences, University of
  California, Berkeley, Rep. UCB/EECS}, 28(13):2009, 2009.

\bibitem{LEE}
C.~A. Lee, S.~D. Gasster, A.~Plaza, C-I. Chang, and B.~Huang.
\newblock Recent developments in high performace computing for remote sensing:
  A review.
\newblock {\em IEEE JSTARS}, 4:508--527, 2011.

\bibitem{ITK}
T.~S. Yoo, M.~J. Ackerman, W.~E. Laorensen, W.~Schroeder, V.~Chalana,
  S.~Aylward, D.~Metaxas, and R.~Whitaker.
\newblock Engineering and algorithm design for an image processing {API}: A
  technical report on {ITK - The Insight Toolkit}.
\newblock In IOS Press~Amsterdam J.~Westwood, ed., editor, {\em Proceedings of
  Medecine Meets Virtual Reality}, 2002.

\bibitem{CHR}
E.~Christophe, J.~Michel, and J.~Inglada.
\newblock Remote sensing processing: From multicore to {GPU}.
\newblock {\em IEEE JSTARS}, 4:643--652, 2011.

\bibitem{GUA2}
Qingfeng Guan, Wen Zeng, Junfang Gong, and Shuo Yun.
\newblock p{RPL} 2.0: improving the parallel raster processing library.
\newblock {\em Transactions in GIS}, 18(S1):25--52, 2014.

\bibitem{QIN2}
Cheng-Zhi Qin, Li-Jun Zhan, A-Xing Zhu, and Cheng-Hu Zhou.
\newblock A strategy for raster-based geocomputation under different parallel
  computing platforms.
\newblock {\em International Journal of Geographical Information Science},
  28(11):2127--2144, 2014.

\bibitem{OTB}
J.~Inglada and E.~Christophe.
\newblock The {O}rfeo {T}oolbox remote sensing image processing software.
\newblock In {\em IEEE International Geoscience and Remote Sensing Symposium
  (IGARSS)}, 2009.

\bibitem{OSA}
Luis Ibáñez and Brad King.
\newblock {ITK}.
\newblock Accessed: 2016-05-24.

\bibitem{WAN}
Lizhe Wang, Yan Ma, Albert~Y Zomaya, Dan Chen, et~al.
\newblock A parallel file system with application-aware data layout policies
  for massive remote sensing image processing in digital earth.
\newblock {\em Parallel and Distributed Systems, IEEE Transactions on},
  26(6):1497--1508, 2015.

\bibitem{THA}
Rajeev Thakur, William Gropp, and Ewing Lusk.
\newblock On implementing mpi-io portably and with high performance.
\newblock In {\em Proceedings of the sixth workshop on I/O in parallel and
  distributed systems}, pages 23--32. ACM, 1999.

\bibitem{QIN}
Cheng-Zhi Qin, Li-Jun Zhan, A~Zhu, et~al.
\newblock How to apply the geospatial data abstraction library ({GDAL})
  properly to parallel geospatial raster {I/O}?
\newblock {\em Transactions in GIS}, 18(6):950--957, 2014.

\bibitem{CHR2}
Emmanuel Christophe and Jordi Inglada.
\newblock Open source remote sensing: Increasing the usability of cutting-edge
  algorithms.
\newblock {\em IEEE Geoscience and Remote Sensing Newsletter}, 35(5):9--15,
  2009.

\bibitem{Note1}
www.equipex-geosud.fr.

\bibitem{Note2}
http://www.theia-land.fr/en.

\bibitem{GIT}
Cresson.
\newblock otbclimpi branch at https://github.com/remicres/otb/tree/otbclimpi.
\newblock Accessed: 2016-05-11.

\bibitem{Note3}
http://www.orfeo-toolbox.org.

\end{thebibliography}

\end{document}